# High pressure crystal growth of the antiperovskite centrosymmetric superconductor SrPt$_3$P

Nikolai D. Zhigadlo[a,b]

[a] *Department of Chemistry and Biochemistry, University of Bern, Freiestrasse 3, 3012 Bern, Switzerland*

[b] *Laboratory for Solid State Physics, ETH Zurich, Otto-Stern-Weg 1, 8093 Zurich, Switzerland*

**Abstract**

Bulk single crystals of SrPt$_3$P have been grown for the first time by a self-flux method at 2 GPa and 1500 °C using the cubic-anvil, high-pressure and high-temperature technique. The grown black crystals were found to have either a plate-like or pillar-like morphology with maximum dimensions of ~ 1 × 0.6 × 0.3 mm$^3$. According to a goniometric study the crystals are elongated in the *ab*-plane with the *c*-axis perpendicular to the plane. The crystal structure was confirmed by X-ray diffraction (*P*4/*nmm*, # 129, *Z* = 2, *a* = *b* = 5.7927(2) Å, *c* = 5.3729(2) Å and V = 180.290(11) Å$^3$). Temperature dependent susceptibility measurements showed a single-phase behaviour and a superconducting transition temperature of 8.6 K. This value for single crystals is slightly higher than that previously reported for polycrystalline SrPt$_3$P. The sharpness of the susceptibility drop ($\Delta T_c$ = 0.07 K) supports a high homogeneity of obtained crystals.

*Keywords*:

A1. Crystal morphology, A1. X-ray diffraction, A2. Single crystals growth, A2. High pressure synthesis

## 1. Introduction

The discovery of superconductivity in Fe-based pnictides [1] has driven the search for superconductivity also in alternative compounds. Among them, Pt-based systems have yielded numerous new intermetallic systems showing myriad structural and bonding features [2]. Evidence of unconventional superconductivity and/or novel physical properties has been reported in various Pt-related compounds, such as SrPtAs [3], $SrPt_2Ge_2$ [4], $SrPtGe_3$ [5], $SrPt_6P_2$ [6], $Ca_2Pt_3Si_5$ [7], MgPtSi [8], $LaPt_5As$ [9], and $Li_2Pt_3B$ [10], as well as in renown strongly-correlated electron systems, such as $UPt_3$ [11] and $CePt_3Si$ [12]. Recently, Takayama *et al*. [13] discovered a new family of ternary platinum phosphide superconductors $APt_3P$ (A = Ca, Sr, and La) with $T_c$'s of 8.4, 6.6, and 1.5 K, respectively. $APt_3P$ compounds adopt a distorted antiperovskite structure, representing the structure of several non-centrosymmetric superconductors such as $LaPt_3Si$ or $CePt_3Si$ (Fig. 1). $SrPt_3P$ seems to be very special among $APt_3P$ compounds, since not only it shows the highest $T_c$ among the 5$d$-electron based superconductors, but it has also a large $2\Delta_0/k_BT_c \sim 5$ value, indicative of strong coupling [13]. The nonlinear magnetic field dependence of Hall resistivity was attributed to the presence of multiple Fermi-surface pockets with two types of carriers. On the other hand, the specific heat data of $SrPt_3P$ were found to be well described within a single band, single *s*-wave gap approach with the zero temperature gap value of $\Delta_0 = 1.85$ meV [13]. Due to these first seemingly contradicting results, $SrPt_3P$ has attracted an increasing interest during the last years [14-24].

By using muon-spin rotation, Khasanov *et al.* [18] studied the magnetic penetration depth ($\lambda$) in $SrPt_3P$ as a function of the applied magnetic field. The established temperature dependence of the penetration depth in the form of $\lambda^{-2}(T)$ again indicated the presence of a single *s*-wave energy gap ($\Delta_0 = 1.58$ meV). At the same time $\lambda$ was found to be strongly field dependent, a characteristic feature of multiband SC systems. The multiband nature of the superconducting state is further supported by the upward curvature of the upper critical field $H_{c2}$(T), just below $T_c$. Such a puzzling behaviour was interpreted to reflect two-band superconductivity with equal gaps but different values for the coherence length [18]. A recent nuclear magnetic resonance (NMR) investigation showed that $SrPt_3P$ behaves as a standard metal in the normal state, being characterized by singlet pairing of the electrons and a gap function with conventional *s*-wave-type symmetry [19].

After the discovery of SrPt$_3$P, several theoretical studies on the pairing mechanism reported partially conflicting results. First-principles calculations performed by H. Chen *et al*. [20] proposed that superconductivity of SrPt$_3$P is caused by the proximity to a dynamic charge- density wave instability and a strong spin-orbital coupling leading to exotic pairing. In contrast, first principles calculations and a Migdal-Eliashberg analysis performed by Subedi *et al*. [21] suggested a conventional phonon-mediated superconductivity. Recently, many efforts have lead to investigate the phonon properties of SrPt$_3$P. Based on the observation of a nonlinear temperature behaviour of the Hall resistivity, Kang *et al*. [22] calculated the electron and phonon band structure of SrPt$_3$P, evidencing that it has low-energy two-dimensional phonons and superconducting charge carriers, which emerge from the $pd\pi$ hybridized bands between Pt and P ions and coupling to low-energy phonon modes. Although Zocco *et al*. [23] provided results for the phonon band structure via inelastic X-ray scattering, further phonon-property analyses are needed for which the availability of single crystal is highly desirable.

*Ab initio* calculation by Nekrasov and Sadovskii [24] confirmed experimental findings concerning the multiple band superconductivity in the SrPt$_3$P compound. In contrast to typical iron pnictides and chalcogenides it was found that the Fermi surface of SrPt$_3$P has a rather three-dimensional shape and consists of two sheets. Due to the two Fermi-surface sheets and to a sizable electron-phonon coupling strength, SrPt$_3$P could exhibit an unusual form of superconductivity: the so-called two-band superconductivity, as was recently suggested by Khasanov *et al*. [18].

As it is well-known from textbooks, the two-band MgB$_2$ superconductor and Fe-based pnictides, multiple sheet Fermi-surface topology may lead to a complicated gap structures with different energy gaps on each sheet [25-29]. The $T_c$ value and the gap ratio in the multiple-band systems are currently being investigated due to a rather complicated interplay of intra- and inter-band couplings. Thus, further progress in exploring the physical properties of SrPt$_3$P superconductor depends crucially on the availability of sufficiently large high quality *single crystals*. Despite extensive experimental efforts spent over the past few years, a number of exciting features, involving electronic and superconducting properties, still remain unsolved or have not been addressed yet. Studies carried out on single crystals can provide significant insight on the interplay among the electron correlation, electron-phonon and spin-orbit coupling in the superconducting phase. This may also pave the route to the understanding of the superconductivity mechanism in other heavy 5*d*-based superconductors.

In this paper, we report details a first successful high-temperature and high-pressure growth of millimetre-sized SrPt$_3$P single crystals. Their structural and superconducting properties are reported below.

## 2. Experimental details

For the growth of SrPt$_3$P single crystals, we applied the cubic-anvil high-pressure and high-temperature (HPHT) technique, developed earlier for growing superconducting MgB$_2$ crystals [30], (Ca,Na)$_2$CuO$_2$Cl$_2$ [31], MgCNi$_3$ [32], $Ln$Fe$Pn$O ($Ln$1111, $Ln$: lanthanide, $Pn$: pnictogen) oxypnictides [33-36], and various other compounds [37]. This apparatus is equipped with a hydraulic oil system comprising a 300 mm semi-cylindrical multi-anvil module. The latter consists of eight outer large steel dies squeezing the smaller six tungsten carbide inner anvils, thus applying an amplified pressure on an experimental cubic cell of $23 \times 23 \times 23$ mm$^3$ in size [Fig. 2(a)]. The edge length of tungsten carbide anvils is 22 mm. The assembly has a single-stage graphite tubular heater (12 mm in inner diameter and 22.3 mm in length) placed in the bore of a pyrophilite cube. The heater is filled with a cylindrical boron nitride (BN) crucible in the center and with pyrophilite tablets on the top and bottom. A current passing through the graphite heater, supplied from outside through the steel parts and two tungsten carbide anvils is used for heating. The temperature is calibrated in advance and is related to the power dissipated in the pressure cell. To avoid overheating the tungsten carbide anvils, a water cooling system is installed.

Stoichiometric amounts of platinum powder (purity 99.99%), red phosphorus powder (purity 99.999%), and strontium pieces (purity 99%+) were thoroughly grounded and pressed into a pellet of approximately 7 mm in diameter and 8 mm of length. The pellet was then placed in a BN container surrounded by a graphite sleeve resistance heater and inserted into a pyrophillite cube. All the work related to the sample preparation and the packing of the high-pressure cell-assembly was performed in a glove box with a protective argon atmosphere. After completing the crystal growth process, the SrPt$_3$P crystals were mechanically extracted from the solidified flux.

The structural properties of SrPt$_3$P single crystals were studied at room temperature on a Bruker X-ray single-crystal diffractometer equipped with a CCD detector and high intensity microfocus x-ray source (Mo K$_\alpha$ radiation). Data reduction and numerical absorption correction were performed by a direct method and refined on $F^2$, employing the SHELX-97 program. The elemental analysis of grown products was performed by means of Energy Dispersive X-ray spectroscopy (EDX, HITACHI S-3000 N). Temperature dependent magnetization measurements were carried out with a Quantum Design Magnetic Property Measurement System (MPMS) XL with the reciprocating sample option installed.

## 3. Results and discussion

The crystallization of inorganic materials under HPHT is an empirical science based on rational trial and error. We carried out a systematic study on the $SrPt_3P$ crystal growth by a self-flux method in order to understand the effects of synthesis parameters on the phase formation and on the morphology of crystals. Figure 2 illustrates schematically the sample cell assembly, the high-pressure synthetic process, and showing pictures of the obtained crystals.

The assembled high-pressure cell was compressed to 2 GPa at room temperature and the optimum growth conditions were tuned by varying the heating temperature, the reaction time and the cooling rate. After optimization, we used a synthesis temperature of about 1500 ºC, found to be optimal for growing sizable $SrPt_3P$ single crystals. Synthesis at lower temperatures (<1400 ºC) resulted in polycrystalline samples. The typical heat-treatment profile for the growth of $SrPt_3P$ single crystals is shown in Fig. 2(b). The BN crucible was heated up to the maximum temperature of ~1500 ºC in 4 h, maintained for 20 h, and then the melt was slowly cooled to 1150 ºC over 40 h and finally reduced to room temperature in 3 h. Figures 2(a) (labels 7 and 8) and 2(c-g) show the results of the high-pressure synthesis. The products inside the crucible were separated into two regions: (i) one at the top end [labelled 7 in Fig. 2(a)] where $SrPt_3P$ single crystals were formed, and (ii) the other at the centre to the bottom end [labelled 8 in Fig. 2(a)], where the rest of the flux had solidified. The original grown lump is shown in Fig. 2(c) where superconducting crystals appeared only in *upper part*. The middle and the bottom parts could be mechanically separated, with the bottom part always showing a rounded shape. Most of the $SrPt_3P$ crystals grown in the upper part were vertically standing, as can be seen from the piece in the Fig. 2(d). Mechanically separated black coloured $SrPt_3P$ crystals were found to exhibit either irregular plate-like [Fig. 2(e-f)] or pillar-like [Fig. 2(g)] shape and they grew in contact with each other. Usually these crystals were free from visible inclusions and cracks reaching a maximum dimensions of ~ $1 \times 0.6 \times 0.3$

mm$^3$. Further examinations of the pillar-like and plate-like shaped crystals have revealed that they belong to the *P*4/*nmm* space group, have composition close 1:3:1 and their $T_c$s are identical.

Currently it is difficult to specify the exact growth mechanism of the SrPt$_3$P crystals. However, from the basic principles of HPHT crystal growth and by taking into account the present results, one can suggest some rational explanations: In our growth attempts we used a stoichiometric composition for the reagents and most probably liquid phosphorus works as a flux which dissolves the other constituents. The growth of single crystals in HPHT apparatus is normally carried out using a temperature gradient. The temperature distribution within each high pressure assembly is determined by the insulating properties of the materials used in a setup, especially those used for the furnace and the sample container. Thermal gradients in the solid medium are the result of heat transport by conduction. In our case, the Sr-Pt-P mixture is loaded into a BN crucible, with the crucible itself being placed in the center of a cylindrical graphite furnace [see for more details Fig. 2 (a)]. The furnace has an axial extent of 22.3 mm in length, which induces heat conduction along the assembly axis and causes a significant temperature gradient in the axial direction. Dotted lines in Fig. 2 (a) schematically show the temperature gradient in our experimental setup, which can be as high as 70 °C when the sample is at 1200 °C. This gradient is one of the important parameters in high-pressure crystal growth. Another driving force for the HPHT growth is the strong influence of the pressure and temperature at which the crystal growth is performed. In a multi-anvil apparatus the quasi hydrostatic pressure results in a homogeneous densification of the sample. However, in reality, at the early stages of the densification process, there is always some pressure that dissipates through the gaskets among the converging anvils. The effective pressure will induce strain, sufficient to generate grain growth during the subsequent sintering. Thus, at high growth temperatures spontaneous nucleation occurs near the initial grains. Each individual high-pressure apparatus shows its own characteristics, which allow one to control the crystal growth at a certain level. For example, the number of the nuclei could be suppressed by making the high pressure furnace short, relative its radius, i.e. by decreasing the axial heat flow. The temperature gradient may be adjusted by changing the thickness of the insulating pyrophilite disks above and below the BN crucible.

In our experimental setup, the cell is heated by passing electrical current through the top and bottom anvils. The average cell temperature increases as more current is passed through the graphite sleeve. However, the top and the bottom ends of the cell are always cooler than its center during the growth process. It is at these cooler places the crystallization usually begins [34]. However, as already mentioned, this is not the case for the SrPt$_3$P crystals. Our growth experiments revealed that SrPt$_3$P crystals *grow only at the top end of the crucible*. We assume that at a sufficiently high supersaturation the spontaneous crystallization of SrPt$_3$P begins at the top of the crucible. At static pressure-temperature conditions, the value of supersaturation depends on the supercooling and mass transport mechanism in the liquid. Further, SrPt$_3$P crystals were produced by directional solidification in the temperature-gradient region of the furnace, where the average temperature decreases gradually. This explanation is supported by the fact that the grown crystals remain vertically oriented in the upper part of the crucible [see Fig. 2 (d)]. Most probably, there the composition of the fluid is different, so that the SrPt$_3$P phase crystallizes only at the top end. Several pieces extracted from the rest of the solidified flux were examined via EDX showing the following composition: "Sr$_{1.0-2.0}$Pt$_{5.3-5.8}$P$_{2.8-3.0}$". Note, that this composition is quite close to that recently reported for the new superconductor SrPt$_6$P$_2$ [6]. As mentioned above, the high-pressure cell is heated from both ends. Thus at high temperature there is sufficient convective motion in the melt. However, as the temperature decreases, the inhomogeneous fluid tends to separate under the influence of gravity: the low-density solution rises, whereas the high-density solution accumulates at the bottom of the crucible. Consequently, the desired SrPt$_3$P crystals appear only in the top part of the crucible, while the rest of the solidified flux yields denser products. Further investigations are under way to clarify the details of the SrPt$_3$P growth mechanism.

Now, let us focus on the structural properties of the grown crystals. Initially, the SrPt$_3$P crystal structure was reported by Takayama *et al*. [13]. To confirm the reported crystal structure we collected data on several SrPt$_3$P pieces, originating from different growth batches. In all cases, a useful single crystal could be found and the refined structural model was basically equivalent to those reported previously for the polycrystalline samples [13].

A full X-ray refinement was performed on a crystal with dimensions of 0.2695 × 0.194 × 0.1327 mm$^3$ ($T_c$ = 8.6 K), using 10472 reflections (of which, 476 unique) in the *k*-space region -11 ≤ $h$ ≤ 11, -11 ≤ $k$ ≤ 11, -10 ≤ $l$ ≤ 10, by a full-matrix least-squares minimization of $F^2$. Structure refinement was performed using the SHELX-97 program [38], in the tetragonal space group *P*4/*nmm*. The crystallographic parameters are summarised in Table 1. The refined atomic positions and bond length are presented in Table 2 and Table 3. The full thermal displacement parameters $U^{ij}$ for all the atoms are reproduced in Table 4. The refinement $R_1$ factor was 7.28% with w$R_2$ =

15.5% including all data. The chemical formula of SrPt$_3$P was confirmed from the fully occupied positions for all sites. As an example, Fig. 3 shows the reconstructed 0*kl*, *h*0*l*, and *hk*0 reciprocal space sections of SrPt$_3$P single crystal measured at room temperature, where no additional phases (impurities, twins, or intergrowth crystals) were detected. The chemical composition was also determined by EDX analysis, which yielded Sr:Pt:P molar ratios close to 1:3:1.

The tetragonal (space group *P*4/*nmm*, # 129, Z = 2) structure of SrPt$_3$P consists of blocks formed by distorted antiperovskite octahedral units Pt$_6$P alternating with atomic sheets of alkaline earth metals. The unit cell parameters $a = b = 5.7927(2)$ Å and $c = 5.3729(2)$ Å are slightly smaller than those reported for polycrystalline samples: $a = b = 5.8094(1)$ Å and $c = 5.3833(2)$ Å [13]. Between Sr layers there are antiperovskite Pt$_6$P octahedral units, where basal Pt1 atoms occupy 4*e* (1/4,1/4,1/2) positions and apical Pt2 occupy 2*c* (0,2.0,0.1407). Phosphorus atoms inside octahedrons also occupy 2*c* position, but with $z = 0.7249$. These octahedrons are not ideal, i.e., the distance from the basal plane to the apical Pt2 atoms are different, while the basal Pt1 atoms form perfect squares. From Fig. 1 it is clear that because of the alternating edge-sharing Pt$_6$P octahedrons the basal Pt1 atoms form a two dimensional square lattice. Pt2 and P atoms make up a Pt2P layer that is geometrically similar to the Fe-As layer in the iron pnictides.

This structure is closely related to that of a heavy fermion superconductor without inversion symmetry, e.g., CePt$_3$Si [12]. In the CePt$_3$P case the unit cell contains only one octahedron and the local lack of inversion symmetry is transferred from the octahedron to the whole solid. However, when the unit cell contains an even number of octahedrons arranged in an anti-polar fashion, the global inversion symmetry is recovered. Such a situation is that actually realized in the SrPt$_3$P compound: the distorted Pt$_6$P octahedral units alternate within the *ab*-planes, forming an anti-polar pattern (Fig. 1). Therefore, SrPt$_3$P contains an inversion center. This is the first example of a P atom encapsulated in an antiperovskite octahedron. As mentioned in Ref. [13], it would be interesting to synthesize both centro- and non-centro-symmetric variants of SrPt$_3$P consisting of electronically equivalent elements. This would allow us to study the effects of the inversion symmetry on superconductivity.

The ZFC and FC magnetic curves for SrPt$_3$P single crystal measured at 3 Oe are shown in Figure 4. The onset of $T_c$ deduced from these curves is 8.6 K, slightly higher than previously reported for polycrystalline samples [13, 15]. For this relatively low field there is almost no coincidence region between the curves. The diamagnetic shielding value (ZFC), after a sample demagnetization correction, approaches 100 % of -1/(4π), whereas the Meissner volume fraction is only 7 % of the ZFC value. This means that even in a field as low as 3 Oe the magnetic flux expulsion is very incomplete, with most of it being pinned and remaining trapped during the cooling process. We can note that the transition temperatures for crystals originating from different parts of the top end were practically identical. On other hand, no signs of a superconducting diamagnetic response were detected down to 1.8 K in pieces extracted from the rest of the solidified flux.

## 4. Conclusion

By using a cubic-anvil high-pressure apparatus single crystals of the SrPt$_3$P superconductor were grown from a stoichiometric melt under a high pressure of 2 GPa at a temperature of 1500 ºC. The crystal structure of SrPt$_3$P was confirmed by single-crystal X-ray diffraction. Temperature dependent magnetization measurements revealed the occurrence of bulk superconductivity showing a sharp transition at 8.6 K. The sufficiently large crystals obtained here open up new possibilities for further investigations on this interesting material.

## Acknowledgements

I would like to acknowledge P. Macchi, T. Shiroka, P. Moll, J. Hulliger and B. Batlogg for comments and helpful discussions.

**References**

[1] Y. Kamihara, T. Watanabe, M. Hirano, H. Hosono, J. Am. Chem. Soc. 130 (2008) 3296.

[2] A. L. Ivanoskii, Platin. Met. Rev. 57 (2013) 87-100.

[3] Y. N. Ishikubo, K. K. Udo, M. N. Dhara, J. Phys. Soc. Jpn. 80 (2011) 1.

[4] H. C. Ku, I. A. Chen, C. H. Huang, C. W. Chen, Y. B. You, M. F. Tai, Y. Y. Hsu, Physica C 493 (2013) 93.

[5] K. Miliyanchuk, F. Kneidinger, C. Blaas-Schenner, D. Reith, R. Podloucky, P. Rogl, T. Khan, L. Salamakha, G. Hilscher, H. Michor, E. Bauer, A. D. Hiller, J. Phys.: Conf. Ser. 273 (2011) 012078.

[6] B. Lv, B. I. Jawdat, Z. Wu, M. Sorolla, M. Gooch, K. Zhao, L. Deng, Y. Y. Xue, B. Lorenz, A. M. Guloy, C. W. Chu, Inorg. Chem. 54 (2015) 1049.

[7] T. Takeuchi, H. Muranaka, R. Settai, T. D. Matsuda, E. Yamamoto, Y. Haga, Y. Onuki, J. Phys. Soc. Jpn. 78 (2009) 085001.

[8] K. Kudo, K. Fujimura, S. Onari, H. Ota, M. Nohara, Phys. Rev. 91 (2015) 174514.

[9] M. Fujioka, M. Ishimaru, T. Shibuya, Y. Kamihara, C. Tabata, H. Amitsuka, A. Miura, M. Tanaka, Y. Takano, H. Kaiju, J. Nishii, J. Amer. Chem. Soc. 138 (2016) 9927-9934.

[10] K.-W. Lee, W. Pickett, Phys. Rev. B 72 (2005) 174505.

[11] G. B. Stewart, Z. Fisk, J. O. Willis, J. L. Smith, Phys. Rev. Lett. 52 (1984) 679.

[12] E. Bauer, G. Hilscher, H. Michor, Ch. Paul, E. W. Scheidt, A. Gribanov, Yu. Seropegin, H. Noël, M. Sigrist, P. Rogl, Phys. Rev. Lett. 92 (2004) 027003.

[13] T. Takayama, K. Kuwano, D. Hirai, Y. Katsura, A. Yamamoto, H. Takagi, Phys. Rev. Lett. 108 (2012) 237001.

[14] R. Szczęśniak, A. P. Durajski, Ł. Herok, Phys. Scr. 89 (2014) 125701.

[15] B. I. Jawdat, B. Lv, X. Zhu, Y. Xue, C.-W. Chu, Phys. Rev. B 91 (2015) 094514.

[16] K. K. Hu, B. Gao, Q. C. Ji, Y. H. Ma, W. Li, X. G. Xu, H. Zhang, G. Mu, F. Q. Huang, C. B. Cai, X. M. Xie, M. H. Jiang, Phys. Rev. B 93 (2016) 214510.

[17] X.-Q. Zhang, Z.-Y. Zheng, Y. Cheng, G.-F. Ji, RSC Adv. 6 (2016) 27060-27067.

[18] R. Khasanov, A. Amato, P. K. Biswas, H. Luetkens, N. D. Zhigadlo, B. Batlogg, Phys. Rev. B 90 (2014) 140507(R).

[19] T. Shiroka, M. Pikulski, N. D. Zhigadlo, B. Batlogg, J. Mesot, H.-R. Ott, Phys. Rev. B 91 (2015) 245143.

[20] H. Chen, X. Xu, C. Cao, J. Dai, Phys. Rev. B 86 (2012) 125116.

[21] A. Subedi, L. Ortenzi, L. Boeri, Phys. Rev. B 87 (2013) 144504.

[22] C.-J. Kang, K.-H. Ahn, K.-W. Lee, B. I. Min, J. Phys. Soc. Jpn. 82 (2013) 052703.

[23] D. A. Zocco, S. Krannich, R. Heid, K.-P. Bohnen, T. Wolf, T. Forrest, A. Bosak, F. Weber, Phys. Rev. B 92 (2015) 220504(R).

[24] I. A. Nekrasov, M. V. Sadovskii, JETP Lett. 96 (2012) 227.

[25] Y. Sassa, M. Månsson, M. Kobayashi, O. Götberg, V. N. Strocov, T. Schmitt, N. D. Zhigadlo, O. Tjernberg, B. Batlogg, Phys. Rev. B 91 (2015) 045114.

[26] T. E. Kuzmicheva, S. A. Kuzmichev, M. G. Mikheev, Ya. G. Ponomarev, S. N. Tchesnokov, V. M. Pudalov, E. P. Khlybov, N. D. Zhigadlo, Phys. Uspekhi 57 (2014) 819.

[27] A. Charnukha, S. Thirupathaiah, V. B. Zabolotnyy, B. Büchner, N. D. Zhigadlo, B. Batlogg, A. N. Yaresko, S. V. Borisenko, Sci. Rep. 5 (2015) 10392.

[28] A. Charnukha, D. V. Evtushinsky, C. E. Matt, N. Xu, M. Shi, B. Büchner, N. D. Zhigadlo, B. Batlogg, S. V. Borisenko, Sci. Rep 5 (2015) 18273.

[29] S. V. Borisenko, D. V. Evtushinsky, Z.-H. Liu, I. Morozov, R. Kappenberger, S. Wurmehl, B. Büchner, A. N. Yaresko, T. K. Kim, M. Hoesch, T. Wolf, N. D. Zhigadlo, Nat. Phys. 12 (2016) 311.

[30] J. Karpinski, N. D. Zhigadlo, S. Katrych, R. Puzniak, K. Rogacki, R. Gonnelli, Physica C 456 (2007) 3-13; N. D. Zhigadlo, S. Katrych, J. Karpinski, B. Batlogg, F. Bernardini, S. Massidda, R. Puzniak, Phys. Rev. B 81 (2010) 054520.

[31] N. D. Zhigadlo, J. Karpinski, Physica C 460 (2007) 372-373; N. D. Zhigadlo, J. Karpinski, S. Weyeneth, R. Khasanov, S. Katrych, P. Wägli, H. Keller, J. Phys.: Conf. Ser. 97 (2008) 012121.

[32] R. T. Gordon, N. D. Zhigadlo, S. Weyeneth, S. Katrych, R. Prozorov, Phys. Rev. B 87 (2013) 094520.

[33] N. D. Zhigadlo, S. Katrych, Z. Bukowski, S. Weyeneth, R. Puzniak, J. Karpinski, J. Phys.: Condens. Matter 20 (2008) 342202.

[34] N. D. Zhigadlo, S. Weyeneth, S. Katrych, P. J. W. Moll, K. Rogacki, S. Bosma, R. Puzniak, J. Karpinski, B. Batlogg, Phys. Rev. B 86 (2012) 214509; N. D. Zhigadlo, J. Cryst. Growth 382 (2013) 75-79.

[35] N. D. Zhigadlo, S. Katrych, S. Weyeneth, R. Puzniak, P. J. W. Moll, Z. Bukowski, J. Karpinski, H. Keller, B. Batlogg, Phys. Rev. B 82 (2010) 064517.

[36] N. D. Zhigadlo, S. Katrych, M. Bendele, P. J. W. Moll, M. Tortello, S, Weyeneth, V. Y. Pomjakushin, J. Kanter, R. Puzniak, Z. Bukowski, H. Keller, R. S. Gonnelli, R. Khasanov, J. Karpinski, B. Batlogg, Phys. Rev. B 84 (2011) 134526.

[37] N. D. Zhigadlo, J. Cryst. Growth 395 (2014) 1; *ibid.*, 402 (2014) 308; C. Müller, N. D. Zhigadlo, A. Kumar, M. A. Baklar, J. Karpinski, P. Smith, T. Kreouzis, N. Stingelin, Macromolecules 44 (2011 1221-1225.

[38] G. M. Sheldrick, (1997) *SHELX-97: Program for Crystal Structure Analysis, University of Göttingen, Göttingen, Germany*.

**Table 1.** Crystal data and structure refinement for SrPt$_3$P.

| | |
|---|---|
| Empirical formula | SrPt$_3$P |
| Formula weight | 703.86 g/mol |
| Temperature | 293(2) K |
| Wavelength | Mo K$_\alpha$ (0.71073 Å) |
| Crystal system | tetragonal |
| Space group | *P4/nmm* |
| Unit cell dimensions | $a$ = 5.7927(2) Å    $\alpha$ = 90° |
| | $b$ = 5.7927(2) Å    $\beta$ = 90° |
| | $c$ = 5.3729(2) Å    $\gamma$ = 90° |
| Cell volume | 180.290(11) Å$^3$ |
| Z | 2 |
| Density (calculated) | 12.966 g/cm$^3$ |
| Reflections collected/unique | 10472/476 |
| Completeness to $\theta$ = 25.242°, % | 100.0 % |
| Data / restraints / parameters | 476 / 0 / 14 |
| Goodness-of-fit on $F^2$ | 1.216 |
| Final *R* indices [$I > 2\sigma(I)$] | $R_1$ = 0.0715, w$R_2$ = 0.1525 |
| *R* indices (all data) | $R_1$ = 0.0728, w$R_2$ = 0.1550 |

**Table 2.** Atomic coordinates ( × $10^4$) and equivalent isotropic displacement parameters (Å$^2$× $10^3$) for SrPt$_3$P.  U(eq) is defined as one third of the trace of the orthogonalized U$^{ij}$ tensor.

| Atom | Wyckoff | x | y | z | U(eq) |
|---|---|---|---|---|---|
| Pt(1) | 4e | 2500 | 2500 | 5000 | 40(1) |
| Pt(2) | 2c | 0 | 5000 | 1407(2) | 38(1) |
| Sr | 2a | 0 | 0 | 0 | 40(1) |
| P | 2c | 0 | 5000 | 7249(14) | 39(1) |

**Table 3.** Bond lengths (Å) for SrPt$_3$P.

| | |
|---|---|
| Pt(1) - P | 2.378(4) |
| Pt(1) - Pt(2) | 2.8146(5) |
| Pt(1) - Pt(1) | 2.86635(10) |
| Pt(1) - Sr | 3.37808(9) |
| Pt(2) - P | 2.234(7) |
| Pt(2) - Sr | 2.9933(2) |
| Sr - P | 3.252(3) |

**Table 4.** Anisotropic displacement parameters (Å$^2$ × $10^3$) for SrPt$_3$P. The anisotropic displacement factor exponent takes the form: $-2\pi^2[\, h^2 a^{*2} U^{11} + ... + 2\, h\, k\, a^* \, b^* \, U^{12}\,]$

| Atom | U$^{11}$ | U$^{22}$ | U$^{33}$ | U$^{23}$ | U$^{13}$ | U$^{12}$ |
|---|---|---|---|---|---|---|
| Pt(1) | 41(1) | 41(1) | 37(1) | 0(1) | 0(1) | 5(1) |
| Pt(2) | 39(1) | 39(1) | 36(1) | 0 | 0 | 0 |
| Sr | 39(1) | 39(1) | 43(1) | 0 | 0 | 0 |
| P | 38(1) | 38(1) | 41(2) | 0 | 0 | 0 |

**Figures and captions**

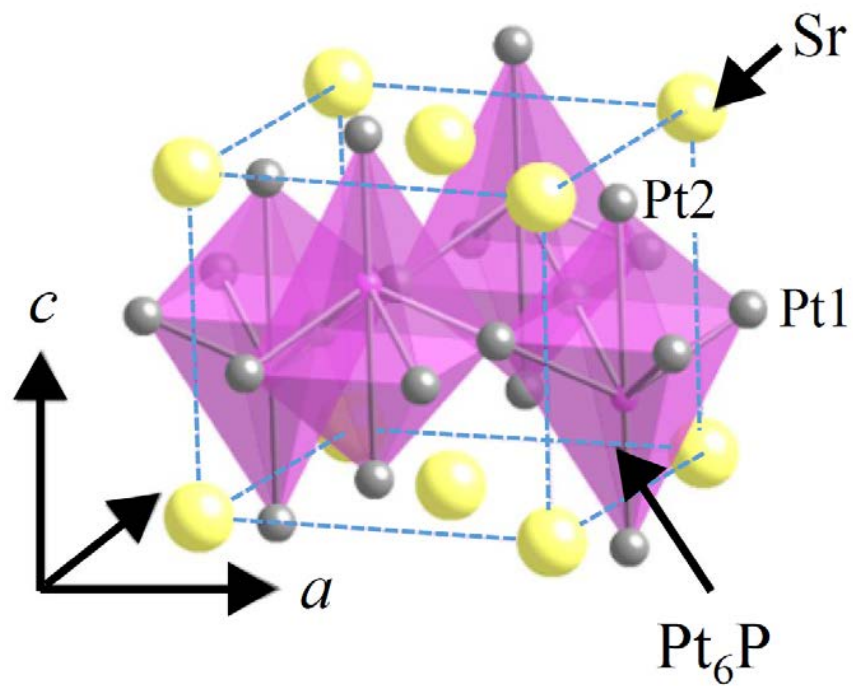

**Figure 1.** (Colour online) Crystal structure of tetragonal (space group *P*4/*nmm*, #129, Z = 2) SrPt$_3$P showing Pt$_6$P octahedrons, arranged in antipolar fashion. Platinum atoms occupy two distinct position: Pt1 and Pt2 (adopted from Ref. 13).

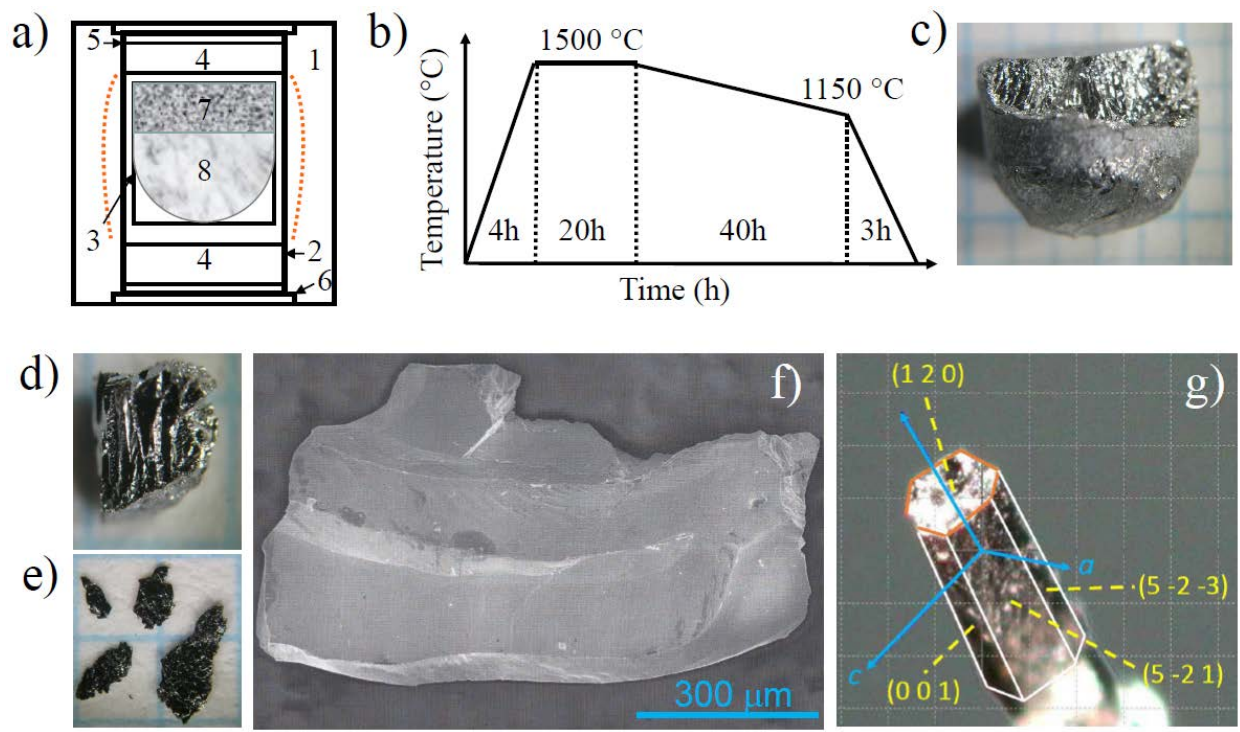

**Figure 2.** (Colour online) Schematic illustration of the sample cell assembly, high-pressure synthesis process, and examples of obtained crystals. (a) The cell assembly for crystal growth: (1) Pyrophyllite cube, (2) graphite sleeve, (3) BN sample crucible, (4) pyrophyllite pellets, (5) graphite disks, (6) stainless disks, (7) SrPt$_3$P crystals grow at the top end, (8) rest of solidified flux. (b) Typical heat treatment profile for the single-crystal growth. (c) Original grown solidified lump. (d) (e) Optical micrographs and (f) scanning electron microscope image mechanically extracted SrPt$_3$P crystals. (g) The growth habit and some faceted planes. The quasi hexagonal morphology may result from twinning in the lateral direction.

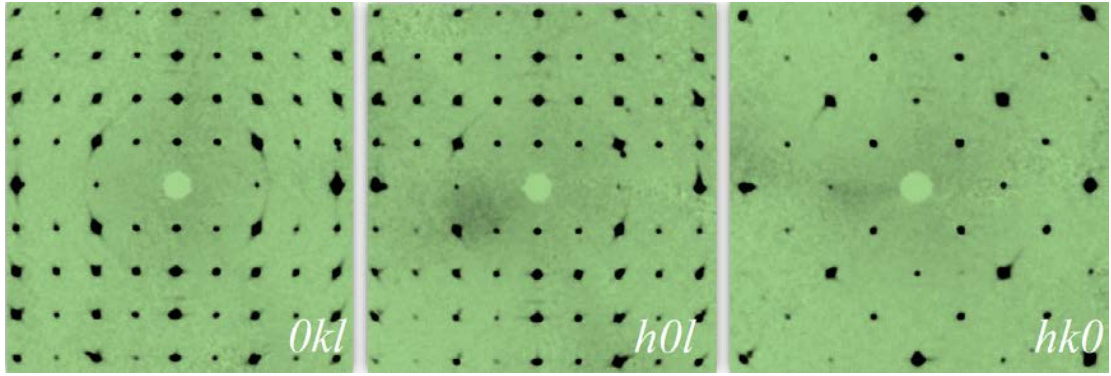

**Figure 3.** (Colour online) The reconstructed *0kl*, *h0l*, and *hk0* reciprocal sections for a single-crystal of SrPt$_3$P. Well-resolved reflections confirm the high quality of grown crystal. No additional phases, impurities, twins or intergrowth crystals were detected.

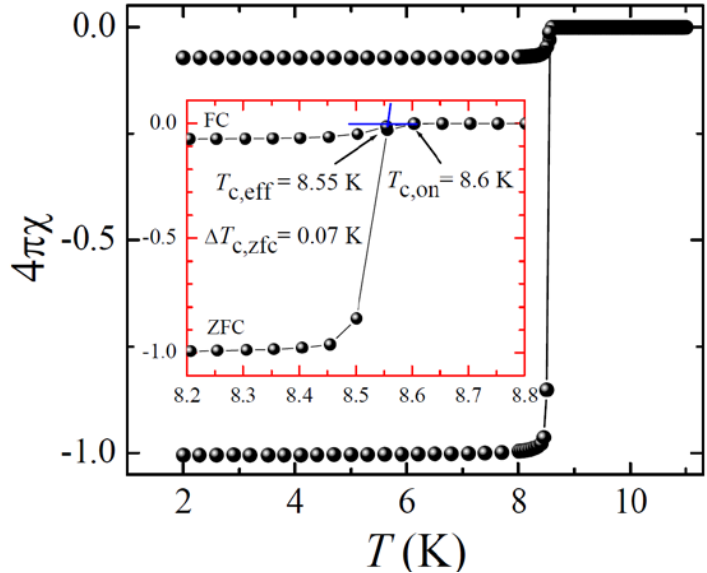

**Figure 4.** Temperature dependence of the magnetic susceptibility for a SrPt$_3$P single crystal. Zero-field-cooled (ZFC) and field-cooled (FC) results are shown. An external magnetic field of 3 Oe was applied. The onset temperature ($T_{c,on}$ = 8.6 K) and the effective transition temperature ($T_{c,eff}$ = 8.55 K) were defined as illustrated in the inset. The transition width $\Delta T_{c,zfc}$ (10 - 90 % criterion) is 0.07 K.

**Highlights**

- SrPt$_3$P crystals have been grown for the first time
- The presence of the temperature gradient plays an important role in the growth
- Plate-like and pillar-like shaped crystals up to ~ $1 \times 0.6 \times 0.3$ mm$^3$
- Sharp superconducting transition at 8.6 K

**Graphical abstract**

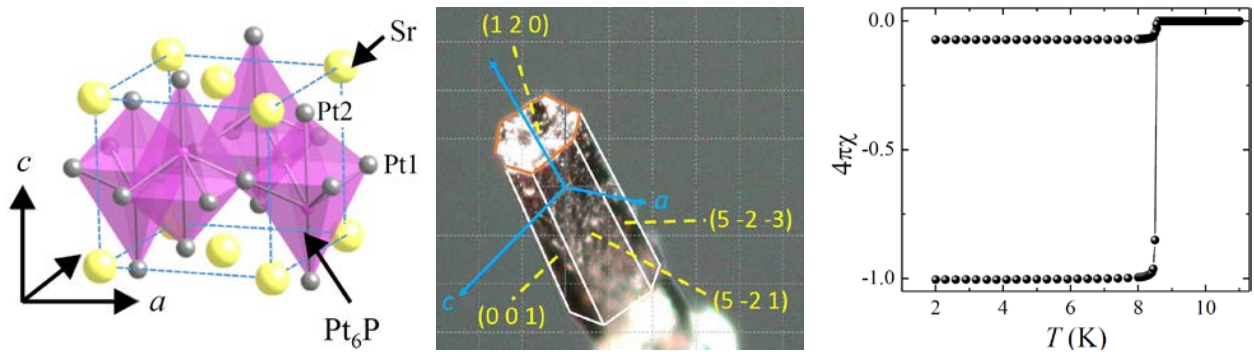